\documentclass[sigconf]{acmart}
\usepackage[utf8]{inputenc}
\usepackage{hyperref}
\usepackage{tikz}
\makeatletter
\usepackage{subcaption}
\usepackage{graphicx}
\usepackage{enumitem}
\newcommand{\result}[1]{}

\newcommand{\done}[1]{}

\usepackage{pifont}

\usepackage{xspace}

\newcommand{\eg}{\textit{e.g.,}~}
\newcommand{\ie}{\textit{i.e.,}~}

\makeatletter
\renewcommand{\paragraph}[1]{\vspace*{0.03in}\noindent{\bf #1.}\hspace{0.25ex \@plus1ex \@minus.2ex}}
\newcommand{\paragraphNoDot}[1]{\vspace*{0.03in}\noindent{\bf #1}\hspace{0.25ex \@plus1ex \@minus.2ex}}
\makeatother

\usepackage[absolute,showboxes]{textpos}
\usepackage[absolute,showboxes]{textpos}
\let\orgautoref\autoref
\renewcommand{\autoref}
{\def\sectionautorefname{\S}%
\def\subsectionautorefname{\S}%
\def\subsubsectionautorefname{\S}%
\orgautoref}
\title[From Files to Streams: Revisiting Web History and Exploring Potentials for Future Prospects]{From Files to Streams: Revisiting Web History \\ and Exploring Potentials for Future Prospects} %

\author{Lucas Vogel}\orcid{0009-0009-5097-4392}
\email{lucas.vogel2@tu-dresden.de}
\affiliation{%
  \institution{TU Dresden}
  \country{Germany}
}

\author{Thomas Springer}\orcid{0000-0003-3221-6677}
\email{thomas.springer@tu-dresden.de}
\affiliation{%
  \institution{TU Dresden}
  \country{Germany}
}

\author{Matthias W\"ahlisch}\orcid{0000-0002-3825-2807}
\email{m.waehlisch@tu-dresden.de}
\affiliation{%
  \institution{TU Dresden}
  \country{Germany}
}

\date{March 2024}

\copyrightyear{2024}
\acmYear{2024}
\setcopyright{rightsretained}
\acmConference[WWW '24 Companion]{Companion Proceedings of the ACM Web Conference 2024}{May 13--17, 2024}{Singapore, Singapore}
\acmBooktitle{Companion Proceedings of the ACM Web Conference 2024 (WWW '24 Companion), May 13--17, 2024, Singapore, Singapore}\acmDOI{10.1145/3589335.3652001}
\acmISBN{979-8-4007-0172-6/24/05}

\makeatletter
\gdef\@copyrightpermission{
  \begin{minipage}{0.3\columnwidth}
   \href{https://creativecommons.org/licenses/by-nc-nd/4.0/}{\includegraphics[width=0.90\textwidth]{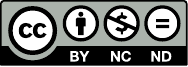}}
  \end{minipage}\hfill
  \begin{minipage}{0.7\columnwidth}
   \href{https://creativecommons.org/licenses/by-nc-nd/4.0/}{This work is licensed under a Creative Commons Attribution-NonCommercial-NoDerivs International 4.0 License.}
  \end{minipage}
  \vspace{5pt}
}
\makeatother

\begin{abstract}

In this paper, we argue that common practices to prepare web pages for delivery
conflict with many efforts to present content with minimal latency, one fundamental goal that pushed changes in the WWW.
To bolster our arguments, we revisit reasons that led to
changes of HTTP and compare them systematically with techniques to
prepare web pages.

\end{abstract}


\begin{document}
\definecolor{boxgray}{rgb}{0.93,0.93,0.93}
\textblockcolor{boxgray}
\setlength{\TPboxrulesize}{0.7pt}
\setlength{\TPHorizModule}{\paperwidth}
\setlength{\TPVertModule}{\paperheight}
\TPMargin{5pt}
\begin{textblock}{0.8}(0.1,0.04)
	\noindent
	\footnotesize
    If you refer to this paper, please cite the peer-reviewed 
publication: L. Vogel, T. Springer, and M. Wählisch. 2024. From Files to 
Streams: Revisiting Web History and Exploring Potentials for Future 
Prospects. In \emph{Companion Proceedings of the ACM Web Conference 2024 
(WWW '24 Companion)}. ACM, New York, USA. 
https://doi.org/10.1145/3589335.3652001

\end{textblock}

\maketitle

\section{Introduction}
\label{sec:intro}
The World Wide Web (WWW) has
changed significantly. Initially, the Web was considered \emph{``a set of
associations, and in a way, the Web is a representation of mankind's
knowledge''}~\cite[Tim Berners Lee]{tiblinterview}. Thirty~years later,
\emph{``the Web is now a gigantic global software
platform''}~\cite[Michael Janiak]{Janiak2023May}. 
Many innovations that led to changes in each part were
driven by creating content more efficiently and delivering content
faster---scalable content dissemination with minimal latency was and
still is critical for successful web applications. 
These changes have
been, so far, mainly considered independently of each other, even though
they could achieve higher performance gains if they are considered
together. One example is the lack of preparing web pages such that
content is segmented into pieces to benefit most from the streaming
capabilities of HTTP/3.

In this paper, we systematically analyze why specific changes have been
made to help our research community identify room for further
improvements. 
We reflect on the different design decisions of
HTTP~(\autoref{sec:httpd}) and common approaches to creating content
(\autoref{sec:content-creation}) based on historical documents and
discussions with key stakeholders. We find that the lack of full
advantages is a rather bad coincidence
(\autoref{sec:historical_overlap}) and derive opportunities for future
improvements~(\autoref{sec:redesign}).

\section{On the History of HTTP}
\label{sec:httpd}

HTTP has evolved significantly since its inception. Originally designed
to deliver linked, mainly text-based documents, it has evolved into an
optimized protocol that enables various applications, including complex
real-time communication. Design decisions of HTTP not only reflect the
needs of emerging web~applications and services but also align with the
state of the larger Internet ecosystem. For instance, when the first web
server implementation was released in
1991~\cite{releaseNotesOfFirstWebServerHttpd}, HTTP was tailored to
transfer linked files, and each web page consisted of relatively few
content pieces. This design choice was inspired by FTP, a popular
standard at the time for downloading files~(RFC 959).
HTTP underwent rapid changes as the Web expanded over the following
years. Now, modern web pages consist of multiple content pieces,
necessitating parallel delivery instead of downloading them
sequentially to form the final page---calling for streams, supported in
more recent HTTP versions.

\paragraph{HTTP/0.9}\label{sec:http-histroy} In 1991, Tim Berners-Lee
designed HTTP ``with simplicity in mind''~\cite{timblhttp09}.
A client requested a web page via a GET request. The original proposal
explicitly states that the server response is HTML \cite{timblhttp09}.
The server then responded with ASCII characters representing the content
of the resource, in this case, a file on the server~\cite{pembertonDesignUrl}. The design lacked error handling. Clients
had to examine the HTML output to determine if the data was correct. 

HTML is based on SGML syntax, as ISO already standardized SGML in 1986. SGML was a file-based format, and to
differentiate both files, the file ending was designed
so it could be renamed from \texttt{.sgml} to \texttt{.html}. From the beginning, HTML allows to
separate structure from layout, differentiating it from SGML's
more generalized and integrated text processing approach~\cite{w3cHistoryOfCSS}. 
In summary, due to prior protocols such as FTP and existing file formats
at the time, the Web was designed to work with files. The lack of
standardization of HTTP and issues, such as the inability to recognize
errors, required further protocol improvements.

\begin{figure*}[ht]
	\includegraphics[width=0.92\textwidth]{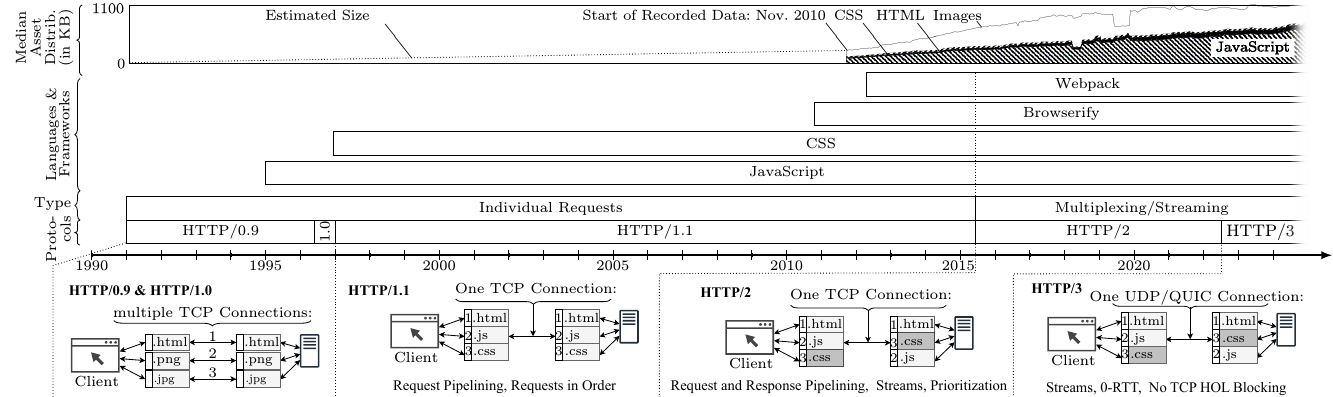}
	\caption{Historical comparison of Web technologies, including protocols
	and content creation and presentation features such as bundlers,
	JavaScript, and CSS. A major change occurred
	when HTTP/2 was released (dotted line).
	Protocol deprecations are not shown. Source for asset distribution
	data: HTTP Archive \cite{httpArchivePageWeight}. Images sizes
	are not stacked, as images are not render-blocking.} 
	\label{fig:timeline}
	\end{figure*}

\paragraph{HTTP/1.0}
Between 1993 and 1997, the Web grew exponentially~\cite{grayWebGrowthSummary}.
This growth motivated clearer documentation and some clean-ups. In May
1996, HTTP/1.0 was published in RFC~1945, an
informational document rather than a standard, collecting the common
usage of existing implementations. In HTTP/1.0~, every
request requires an individual TCP handshake, which is closed after the
answer is transferred, see \autoref{fig:timeline}. The
increase in web pages led to a stronger focus on the presentation layer
(see \autoref{sec:content-creation}). HTTP now supported Content-Types, allowing for different media to be transferred since
the type of content was specified explicitly. This shift in perspective
could have marked a turning point for the Web, enabling truly mixed
content in a single request. Until today, a significant number of media
types have been registered, maintained by
IANA. However, potentially because text-based
implementations already existed, and with document formats such as
images or CSS files, the file-based structure prevailed. 

\paragraph{HTTP/1.1}
HTTP/1.1, specified in 1997 in RFC 2068 and later RFC~2616, addressed a fundamental performance problem of
HTTP/1.0 by introducing two features. First, instead of using individual
TCP requests for every resource, which slowed down the transmission
because resources (\eg images) required a handshake as well, HTTP/1.1
allowed reusing one TCP connection for multiple resources (see
\autoref{fig:timeline}) based on the Keep-Alive mechanism. Second,
HTTP/1.1 introduced request pipelining, where multiple requests could be
sent via a single TCP connection, \ie a client can send
requests in order without waiting for the responses. 

Session reuse and request pipelining directly addressed protocol-level
needs to present content faster. Using a single connection for multiple
types of content, \eg CSS or HTML, improved loading times, for
example, by preventing the execution of TCP slow start 
multiple times. However, pipelining was deactivated by default in a
majority of modern HTTP clients~\cite{mozillaConnectionManagementHTTP1x}
due to multiple problems such as buggy proxies and the complexity of
implementation.

Starting in July 1997, the newly formed HTTP-NG Working Group proposed
several ideas for a new generation of HTTP \cite{httpNGWorkingGroup}.
They used the concept of multiplexing data into one single stream. This
would allow for faster transfer, as HTTP pipelining must wait
until a response is fully finished before new data can be sent
\cite{mozillaConnectionManagementHTTP1x}. In 1999, the outcome was
transferred to the IETF~\cite{httpNGWorkingGroup}.

\paragraph{HTTP/2}
In 2009, Google announced SPDY, a project aiming to improve web page
loading speed by minimizing latency \cite{SPDYannouncement}. This
approach also utilized multiplexing, with Google claiming up to a 55\%
reduction in page loading time over SSL. Some of
the inspiration was drawn from HTTP-NG. As SPDY
gained more traction, the HTTP Working Group specified the HTTP/2
protocol in 2012, inspired by SPDY \cite{httpgwRecharterSpdy}. In May 2015, HTTP/2 was published as RFC 7540. Some major technical improvements include data
compression, request prioritization, server push, and, most notably,
multiplexing requests. They allow for transferring binary data in
parallel in a different order based on prioritization, as shown in
\autoref{fig:timeline}. Multiplexing is a significant step because it
changes the original design of a file-based protocol to streams. On a
protocol level, streams were the new way of transferring data on the
Web. Despite speed improvements, streams were primarily used for
transferring files, such as CSS, HTML, or JavaScript.

\paragraph{HTTP/3}
Before HTTP/3, TCP was the preferred protocol for transmitting HTTP. However, TCP acknowledges every packet,
causing delays if a packet is lost, as all streams are stopped. This
issue is known as HOL-blocking (head-of-line blocking). 
Google started working on QUIC in 2012, an alternative to TCP, TLS, and
HTTP/2. QUIC is based on UDP and implements a larger
part of the stack, allowing for faster protocol updates. 

In 2018, the IETF decided that ``HTTP/QUIC'' should be named
HTTP/3~\cite{ietf103Proceedings-httpsbis-minutes,ietf103Proceedings-httpsbis-bishop}.
Later, QUIC was published as RFC~9000 in~2021
and HTTP/3 as RFC~9114 in 2022.
HTTP/3 retains the stream-based approach but introduces new features,
such as per-stream flow control.
With the replacement of HTTP/2 by HTTP/3 and TLS and TCP by QUIC,
HTTP has become a specialized protocol designed to transfer client
content based on low-latency connection setup and better system handling of lost
packets. 

In summary, HTTP has changed over the last 30~years.  Web content, however, has also changed, partly independently of HTTP, partly to cope with its limitations. We discuss these changes~next.

\section{Web Content Creation History}
\label{sec:content-creation}

The limitations of HTTP also influenced the creation of web content. In
the context of this paper, the term ``content'' refers to code that
is sent to the user and processed there to be displayed as a web page.
The design choices for preparing code are sparsely documented compared
to protocol~changes. 

\paragraph{Beginning of Scripts and Styles}
In 1990, the Web Browser written by Tim Berners-Lee already included an
editor for users to modify content~\cite{firstWebbrowserTimBernersLee}. Subsequent browsers,
such as the Line Mode Browser or Viola, did not have a built-in editor. Since then,
the creation and consumption of web content have diverged. 

In October~1994,
H\aa kon Wium Lie proposed Cascading HTML style sheets (CHSS)
just three days before Netscape's announcement \cite{cssFirstProposal,
w3cHistoryOfCSS}. After collaborating with Bert Bos and
significant debates at WWW conferences, the first version of
Cascading Style Sheets (CSS1) was published in 1996
\cite{w3cHistoryOfCSS, cssLevel1Start1996}.
Meanwhile, JavaScript emerged in 1995, when Netscape added scripting
into their browser, realizing the need for a more dynamic
Web. This
involved two strategies. First, Brendan Eich was hired to work on a
Scheme at Netscape \cite{JavaScriptBrendanEichPopularity}. Secondly,
Netscape collaborated with Sun (now part of Oracle) to provide Java
Applets. After
Eich wrote a prototype in 1995, it was named ``LiveScript'' by Netscape
marketing but later renamed to JavaScript \cite{birthOfJavaScript}. JavaScript was, and still is, the prominent
scripting language of the Web. 

\paragraph{Early Development of Websites with JavaScript}
The release of the first DOM
specification in 1998 set the starting
point for extended JavaScript development, enabling dynamic web
applications. According to various informal sources, developing
JavaScript with multiple files (separated for an improved developer
experience) often used concatenation to produce one output file, which
could then be sent to the user \cite{historyOfModulesAndBundling, stateOfBundlersAndBuildTools}. These were
individual, large global scripts \cite{historyOfModulesAndBundling}. One
reason for this combination of code, such as JavaScript or CSS, was
loading time optimization. Since HTTP/1.0 required a new TCP connection
for every request, combined with the various issues with HTTP/1.1
implementations meant that combining resources could generally improve
performance \cite{mozillaConnectionManagementHTTP1x}. Additionally,
simultaneous connections to a single endpoint were generally limited to
six TCP connections per individual origin (a host name and port) by
browsers supporting HTTP/1.1 \cite{Grigorik2015Nov}.
Consequently, combining resources was necessary, or Domain Sharding was
required where multiple subdomains are used to host a larger number of
files \cite{Grigorik2015Nov}. Despite these restrictions, JavaScript and
CSS grew significantly in popularity, partially due to the dotcom boom
at the time. Resource concatenation was therefore
utilized to improve performance. 

\setlength{\abovecaptionskip}{5pt plus 3pt minus 2pt}
	
\paragraph{The Dynamic Web}
Since 1993, CGI (RFC~3875) has been a standardized way of adding dynamic
functionality to web pages by running scripts on servers. However,
security concerns emerged~\cite{w-ccp-98}.
Client-side code executed in the browser can avoid this issue.
Therefore, tools such as Java Applets, Adobe Flash, and Shockwave gained
popularity when the Netscape Plug-in API was released in~1996.

In 2005, Garrett introduced ``Ajax''~\cite{ajax}, a new way of loading
content asynchronously. 
Before Ajax, presenting updated content under the same URL often required a
reload of the web page.
Given the connection overhead of HTTP/1.1, a complete reload of a page
slowed down the interaction speed \cite{ajax}. The Ajax concept paved the way for ``Web
2.0.'' \cite{blank2012participatory}. 
Partially to simplify the development and implementation of Ajax, Resig
introduced jQuery~\cite{r-j-06} in 2006, which since then has been a popular library
based on JavaScript.
jQuery is still used by 77\% of all websites \cite{jsLibrariesUsageStatistics}. 

With the increasing popularity of jQuery, prior approaches were less
deployed. Java Applets were deprecated in
2017, and Shockwave and Adobe Flash
in 2019 and 2020.
Libraries such as jQuery, Bootstrap, and Underscore simplify the
developer experience and their success
demonstrates that the broad adoption of new technologies depends on an
easy-to-use development.
The downside is that such libraries can increase
loading times as they must be loaded completely before use. By
default, this is render-blocking, preventing the browser from
displaying the page.
Furthermore, the use of libraries and an increasing amount of JavaScript
also result in web pages continuously growing in size. Even though full
page reloads are no longer necessary, the overhead of loading large
render-blocking libraries can still impact loading speed, even with
improved transfer speeds of HTTP/2 or~HTTP/3.

\paragraph{Start of Node.js and Bundles}
In order to organize the increasing amount of JavaScript code,
developers separated code into individual files, which were then
concatenated and used as global scripts~\cite{historyOfModulesAndBundling, stateOfBundlersAndBuildTools}. 
That changed when Node.js was introduced
by Ryan Dahl in 2009 \cite{nodejsFirstNameRyanDahl}. Node.js allowed
JavaScript to run on a server and shipped with a package manager called
\textit{npm}. Npm allowed developers to install other packages and
import (``require'') code when needed.
In contrast, JavaScript did not have a native module
system until 2015 \cite{historyOfModulesAndBundling,
nativeModulesJavaScript}. This module system of Node.js enabled
significantly larger code bases. Node.js, however, was not intended to
be used in a browser. 
This changed when Browserify emerged in 2010 and became popular in 2013
\cite{googleTrendsBrowserifyWebpackReactAngular}.
Browserify allows using the CommonJS ``require'' syntax, allowing
developers to use and require npm packages while creating websites
\cite{historyOfModulesAndBundling}. However, browsers do not understand
the ``require'' keyword, so Browserify transforms all code into a
browser-compatible version. All dependencies are resolved and combined
into a single file, called a
\emph{bundle}~\cite{historyOfModulesAndBundling}. Even though the
primary goal of Browserify was different, it is often cited as being the
first influential bundler \cite{historyOfModulesAndBundling, stateOfBundlersAndBuildTools}. In summary,
bundling tools allow combining multiple code files. Before reducing the
overhead of individual requests started in HTTP/1.1, bundling
improved loading, as fewer requests were required to fetch 
necessary data for page load while also improving developer~experience.

\paragraph{Webpack}
While Browserify could compile and bundle code, the purpose-built
compiler and bundler Webpack~\cite{webpackWebsite}, started in 2012, gained
popularity in $\approx$2015 \cite{historyOfModulesAndBundling,
googleTrendsBrowserifyWebpackReactAngular}. In
contrast to Browserify, it was also designed to bundle other resources,
such as CSS or images. This enabled the emergence
of popular modern frontend frameworks such as React or Angular, both of
which use Webpack. Especially with React gaining popularity, Webpack
became more popular than Browserify
\cite{googleTrendsBrowserifyWebpackReactAngular}. Splitting code is
possible with Webpack as a manual opt-in feature but is disabled by
default~\cite{webpackCodeSplitting}. Webpack was designed for large
projects, but for
significantly large projects it is not advised to load the bundle as one large file. 
Therefore, popular frontend frameworks and bundlers like
Webpack will produce large, individual files that, by default, are
render-blocking~\cite{googleTrendsBrowserifyWebpackReactAngular}.

\section{Historical overlap of protocol changes and content creation}
\label{sec:historical_overlap} 

Contextualizing HTTP and content creation into historical context, it
becomes evident that changes to HTTP were misaligned in time with
changes to languages and content creation frameworks, as shown in
\autoref{fig:timeline}. Until 2015, HTTP/1.1 (and predecessors) on top
of TCP led to an environment where larger but fewer requests improved
performance due to the recurring overhead of connection establishment
per request. To overcome the performance drawbacks of HTTP and TCP,
bundlers such as Webpack gained popularity because using large
individual files (bundles) fit the needs of developers and reduced
performance penalties \cite{historyOfModulesAndBundling,
nativeModulesJavaScript, googleTrendsBrowserifyWebpackReactAngular}. The
major drawback of bundling is code efficiency. If a project uses
JavaScript or CSS frameworks, the entire code base of the framework is
usually fully included in a bundle. This increases the total size of a
web page and, if loaded in the default render-blocking way, slows down
the loading time of web pages. A recent study revealed that the majority
of the most popular web pages still have this issue
\cite{vogel2022depth}.

In 2015, with the advent of HTTP/2, the system changed. Now, a single
connection can multiplex various resources with different
priorities, allowing for the asynchronous transfer of smaller files. This change is in direct contrast to the concept of
bundling and the emergence of libraries, such as jQuery. Hofman, a Web
performance expert, called bundling an ``anti-pattern'' in HTTP/2~\cite{hofmanBundlingAntiPattern}. In principle, the content of a web
page could be adapted again to benefit from the features of HTTP/2.

\vspace{-0.6cm}
\section{New content structure and HTTP/3}
\label{sec:redesign}

\paragraph{Basic Concept}
While bundling uses the central concept of combining resources, the next
generation of web content should do the opposite: split the content as
much as possible~\cite{hofmanBundlingAntiPattern}. 
If code is split into small, individual pieces, then it can
be loaded asynchronously on demand. If planned correctly, the
transferred code only contains data required for rendering, which can
significantly improve loading times. 
Splitting is relevant for CSS and
JavaScript, as both can be loaded externally and both can be render-blocking. 
The next generation of code processors should take advantage of the prevailing streaming technology available in
HTTP/2 and HTTP/3. This moves loading from an all-or-nothing approach to
the behavior of loading a web page as an ongoing process over time. Such
a concept has been studied~\cite{jahromi2020beyond}, stating that users do not wait for a
complete result after an interaction. Streaming can be used as a major
advantage, presenting the user with a continuously rendered version as
the page loads to enable subsequent actions. This, combined with the
maximum splitting approach, has the additional positive effect that
a First Contentful Paint (FCP) of a page can be fast, even in challenging
network conditions. Three main challenges remain to provide a similar, fully automated system like bundlers:

\paragraph{Challenge 1: Content usage detection} The first step is the
automatic generation of information about the type of code used and where
to split the code. For CSS, techniques such as Critical~\cite{criticalCSS} exist but are limited to
	the content above-the-fold. 
	Other~solutions~include the Essential framework, which addresses the issues of Critical \cite{vogel2023speed}. For JavaScript, 
	approaches like tree shaking are available and already in use by
	bundlers today \cite{treeShaking}. They remove dead code but identify
	dead code only on file or function level and
	are bound to specific frameworks. One approach to improve the situation
	is resumeability, where
	JavaScript code is loaded on demand without breaking the web page.
	This approach requires less developer effort and is
	framework-independent.

\paragraph{Challenge 2: Content usage location and order} After detecting and splitting the necessary code, it needs to be ordered and
	interleaved so that only neighboring pieces of HTML, CSS,
	and JavaScript depend on each other and, thus, only minimally block
	rendering. For CSS, this involves matching selectors with the
	DOM of a document. For JavaScript, this is still an open
	challenge, but promising automation approaches have been proposed~\cite{Fawkes}.

\paragraph{Challenge 3: Streaming a web page} Lastly, the processed data
	needs to be transferred by utilizing the streaming capability of
	HTTP/2 and HTTP/3. 
	Streaming of web page chunks~\cite{turbo} as well as
	complete web pages \cite{vogel2023streaming} is feasible.

It is worth noting that streaming is just one possible solution
but---looking at the continuous trend of ever-increasing web pages---it
appears a promising approach \cite{httpArchivePageWeight} to reduce loading
times~\cite{vogel2023streaming}.

\label{lastpage}

\bibliographystyle{ACM-Reference-Format}
\bibliography{bibliography,rfcs}
\end{document}